\documentclass[%
superscriptaddress,
 amsmath,amssymb,
 aps,
 pra,
 10pt,
]{revtex4-2}

\usepackage{tikz}
\usepackage[english]{babel}
\usetikzlibrary{trees}
\usetikzlibrary{decorations.pathmorphing}
\usetikzlibrary{decorations.markings}
\usepackage{tikz-cd}
\usetikzlibrary{arrows,arrows.meta}

\tikzset{
    photon/.style={decorate, decoration={snake,segment length=1.5mm}, draw=black},
    electron/.style={draw=black, postaction={decorate},
        decoration={markings,mark=at position .55 with {\arrow[draw=black]{>}}}},
    gluon/.style={decorate, draw=magenta,
        decoration={coil,amplitude=4pt, segment length=5pt}},
    boundelectron/.style={thick, double}
}

\usepackage{hyperref}
\hypersetup{
    colorlinks=true,
    linkcolor=black,
    filecolor=black,      
    urlcolor=blue,
    citecolor=black,
}

\usepackage{graphicx}
\usepackage{siunitx}
\usepackage{dcolumn}
\newcolumntype{.}{D{.}{.}{8}}
\usepackage{bm}
\usepackage{rotating}
\usepackage{comment}
\usepackage{empheq}
\usepackage[T1]{fontenc}
\usepackage[utf8]{inputenc}
\usepackage{physics}
\usepackage{amsmath, amssymb, amsthm}
\usepackage{tabularx,booktabs,array,dcolumn}
\usepackage{setspace}
\usepackage{vmargin}
\usepackage{xcolor}
\usepackage{graphicx,psfrag,subfigure}

\usepackage{float}
\usepackage[all]{xy}
\usepackage{verbatim}
\usepackage{dsfont}
\usepackage{empheq}

\usepackage{slashed}
\usepackage{multirow}

\newcommand{\bos}[1]{\boldsymbol{#1}}
\newcommand{\mr}[1]{\mathrm{#1}}

\def\iim{\mr{i}}
\def\eem{\mr{e}}
\def\dd{\mr{d}}

\def\nb{n_\text{b}}

\def\bR{\bos{R}}

\def\bp{\bos{p}}

\def\bx{\bos{x}}

\def\balpha{\boldsymbol{\alpha}}
\def\bsigma{\boldsymbol{\sigma}}

\def\he2p{He$_2^+$}
\def\h2p{H$_2^+$}

\newcolumntype{d}[1]{D{.}{.}{#1}}

\begin{document}

\title{Two-center Dirac equation in a Gaussian basis set}
\author{D\'avid Ferenc} 
\email{david.ferenc@irsamc.ups-tlse.fr}
\affiliation{Laboratoire de Chimie et Physique Quantique UMR 5626 CNRS---Université de Toulouse, 118 Route de Narbonne, F-31062 Toulouse, France}

\date{\today}

\begin{abstract}
\noindent
A method for solving the two-center Dirac equation using a Gaussian-product basis set tailored to the axial symmetry of diatomic systems is presented. For the ground state of H$_2^+$ at an internuclear distance $R=2.0~a_0$, we obtain a relativistic energy with relative precision at the $10^{-18}$ level. The properties of the basis set allow for an analytic Fourier transform of the wave function. As a first application we evaluate, for the first time, the zero-potential contribution to the one-loop self-energy for H$_2^+$ in Feynman and Coulomb gauges. Although this contribution is not separately gauge invariant, its calculation provides a proof of principle for the use of the present framework in prospective QED calculations.

\end{abstract}
\maketitle
\section{Introduction}
\noindent
While the nonrelativistic two-center Coulomb problem played a fundamental role in the understanding of chemical bonding in the early years of quantum mechanics~\cite{HeitlerLondon1927,Burrau1927,Wilson28_1,Wilson28_2,Teller1930,Hylleraas1931}, its relativistic counterpart is still an active area of research today~\cite{NoKoKa22,KullieSchiller22,NogueiraKarr23,Kullie24}. The first precise numerical solution to the Dirac equation with a two-center potential was achieved in Refs.~\cite{MuRaGr73,MullerGreiner76}, using a generalized Hylleraas basis set in prolate spheroidal coordinates. This work was motivated by the theoretical description of heavy-ion collision experiments, where relativistic effects play a significant role and cannot be treated perturbatively. 

More recently, applications have shifted to the lightest molecular systems, the hydrogen molecular ion H$_2^+$ and its isotopologues. These are among the most accurately studied systems, both theoretically~\cite{KoHiKa17,KorobovKarr21,KuNoKa25} and experimentally~\cite{Patra20,AlGiCoKoSc20,KoAlHaGiKoSc21,AlKoGiSc23,ScAlSc24,Alighanbari25,HoScScStStFeHoKi25,Konig26}. High-accuracy theoretical predictions require the precise evaluation of relativistic and QED corrections to the energy levels. Such calculations are typically carried out in the nonrelativistic QED (NRQED) framework, where the corrections are obtained from an expansion both in the fine-structure constant $\alpha$ and in powers of the binding parameter $Z\alpha$, where $Z$ is the nuclear charge \cite{KorobovKarr21}. This method is well suited for light, few-particle systems, yet the higher-order corrections become increasingly complex to evaluate. Currently, the accuracy of theoretical predictions for spin-averaged transition frequencies in hydrogen molecular ions is limited by uncalculated higher-order contributions to the one-loop self-energy~\cite{KorobovKarr21}. For light, hydrogen-like atoms, these terms were obtained from the computation of the self-energy to all orders in $Z\alpha$~\cite{JeMoSo99,JeMoSo01,YeHaKe25}. For molecular ions, however, the numerical construction of the Green's function for an electron bound in the two-center Coulomb potential is necessary, which requires a precise representation of the spectrum of the Dirac--Coulomb operator. While all-order self-energy computations have previously been carried out for heavy diatomic quasi-molecules using a $B$-spline method combined with the multipole expansion of the two-center potential~\cite{ArSu15,ArSuYe22}, the accuracy is limited by the truncation of the expansion, and the computations become increasingly challenging for larger nuclear distances. Therefore, for high-precision computations, an approach that treats the cylindrical symmetry of the system explicitly is preferred.

Several methods have been developed over the years for the solution of the two-center Dirac equation. Computations have been carried out using linear combination of atomic orbitals constructed from Gaussian-~\cite{MaLiRo80,MarkSchwarz82,MarkBecker87,FrankeKutzelnigg92,ParpiaMohanty95,Franke97,AlexanderColdwell99,Rutkowski99} and Slater-type orbitals~\cite{PavlikBlinder67,LaJohnTalman92,LaJohnTalman98,AlexanderColdwell99}. Furthermore, various finite-basis methods adapted to curvilinear coordinates~\cite{IsNaNa08,IsNaNa12,NoKoKa22,NogueiraKarr23}, finite-element methods~\cite{SchulzeKolb85,YaHeKo91,KullieKolb2001,KuKoRu04,ArSuInPlSt10,FiLoBa12,KullieSchiller22,ZaToDaLoDeMa24,SoAnDaGlKo24,Kullie24}, and other numerical finite-difference methods were also applied~\cite{LaaksonenGrant84,SuPyLa87,Sundholm1994,TuKoShDeHaKoPlSt10,TupitsynMironova14,KoKe22}.
Recently, a precise finite-element solution was used for the computation of relativistic corrections to the bound-electron $g$-factor, contributing to the most accurate evaluation of the quantity to date~\cite{KuNoKa25}. The most precise relativistic energies for H$_2^+$ were obtained using exponential-type basis sets, reaching up to 32 significant digits~\cite{NogueiraKarr23}. While the precision of these computations is outstanding, the method relies on computationally expensive multiprecision arithmetic due to linear dependencies and large numerical cancellations during the evaluation of matrix elements. 

Here, we propose the use of Gaussian-product basis sets. Their analytic properties allow for numerically robust evaluation of multicenter integrals, operator expectation values, and analytic Fourier transforms, which will facilitate the computation of QED corrections for diatomic molecules in the future. The use of Gaussian basis sets has already been explored for the evaluation of QED corrections for simple atomic systems in recent years. The Wichmann--Kroll contribution to vacuum polarization has been computed with Gaussians for highly charged hydrogen-like ions independently by several groups~\cite{SaSa23,IvBaGlVo24,ZoZeOr26,BeSaSa26}. It was recently demonstrated that Gaussian basis sets are also well suited for the computation of the bound-electron self-energy correction in hydrogen-like ions~\cite{FeSaSa25}. More recently, a similar study was carried out with exponential-type basis sets~\cite{NoSaKa26}, also with applications to atoms with low nuclear charge, adopting the approach of Ref.~\cite{YeHaKe25}. A comparison of the self-energy computations performed with the two types of basis sets in Ref.~\cite{NoSaKa26} shows no clear improvement in accuracy using exponential-type functions instead of Gaussian basis. Although the basis set parameters were not optimized in either computation, using twice as many exponential basis functions led only to a modest reduction in the uncertainty of the self-energy correction compared to the results of Ref.~\cite{FeSaSa25}. These observations, together with the advantageous mathematical properties of Gaussian functions, motivate the development of a Gaussian basis set method for the solution of the two-center Dirac equation that may be applied to all-order self-energy computations.

In the present work, we do not aim to surpass the highest-precision two-center Dirac equation computations in terms of the number of significant digits. Instead, we aim to develop a numerically robust and compact representation of the relativistic wave functions that preserves most of the analytic properties of radial Gaussian basis sets, which have already been successfully used in atomic self-energy computations. We propose a Gaussian-product Ansatz tailored to the axial symmetry of diatomic molecules and show that accurate relativistic energies can be obtained with moderate-sized basis sets. An important feature of the Ansatz is its analytic Fourier transform. As a first application, we evaluate the zero-potential contribution to the one-loop self-energy. Although this is only a partial contribution to the complete self-energy correction, it provides a direct test of the momentum-space wave function. The accuracy obtained here is sufficient for the subsequent calculations of momentum-space QED contributions. We use atomic units unless otherwise stated. The value of the inverse fine-structure constant is set to the CODATA18 value for comparison with previous work, $\alpha^{-1} = 137.035\, 999\, 084$~\cite{codata18}.

\section{Method}
\subsection{Coordinate-space wave function}
The Dirac equation is
\begin{align}
    h_\text{D} \psi_{a}(\bx) = E_{a}\psi_{a}(\bx)  \, ,
    \label{eq:direq}
\end{align}
with the Dirac--Coulomb Hamiltonian 
\begin{align}
    h_\text{D}
    = 
    c\balpha\cdot \bp + (\beta -1) mc^2 + V
    =
    \begin{pmatrix}
    V & c\boldsymbol{\sigma}\cdot\boldsymbol{p}\\
    c\boldsymbol{\sigma}\cdot\boldsymbol{p} & V-2mc^{2}
    \end{pmatrix}
    \, ,
\end{align}
where $mc^2$ has been subtracted to align with the nonrelativistic energy scale. The Coulomb potential of two clamped nuclei with nuclear charges $Z_1$ and $Z_2$ is
\begin{align}
    V(\boldsymbol{x})
    = 
    V_{1}(\boldsymbol{x})+V_{2}(\boldsymbol{x})
    =
    -\frac{Z_{1}}{|\boldsymbol{x}-\boldsymbol{R}_{1}|}
    -\frac{Z_{2}}{|\boldsymbol{x}-\boldsymbol{R}_{2}|}
    \, ,
\end{align}
where $\bR_1$ and $\bR_2$ are the coordinates of the nuclei. The axial symmetry allows for the reduction of the equations to two dimensions, which we carry out in cylindrical coordinates $\{r,z,\phi\}$. The transformation from cylindrical to Cartesian coordinates is given by
\begin{align}
\begin{split}
    x &=  r\cos\phi\\
    y &=  r\sin\phi\\
    z &=  z \, .
\end{split} 
\end{align}
The inverse transformation is 
\begin{align}
    \begin{split}
    r    &= \sqrt{x^2+y^2} \\
    \phi &= 
        \begin{cases}
             \arccos\left(\frac{x}{r}\right) & \text{if}\  y\geq 0  \\
            -\arccos\left(\frac{x}{r}\right) & \text{if}\  y<0 \, .
        \end{cases}
    \end{split} 
\end{align}
The azimuthal angle $\phi$ remains undefined for $r=0$. The necessary differential operators are easily expressed in this coordinate system. The Laplacian
\begin{align}
    \Delta  
    &=  
    \frac{\partial^{2}}{\partial r^{2}}
    +
    \frac{1}{r}\frac{\partial }{\partial r}
    +
    \frac{1}{r^{2}}\frac{\partial^{2}}{\partial\phi^{2}}
    +
    \frac{\partial^{2}}{\partial z^{2}}
    \, ,
\end{align}
and the components of the $\bsigma\cdot\bp$ operator are
\begin{align}
    \bsigma\cdot\bp
    &=
    \begin{pmatrix}
    p_z & p_x - \iim p_y \\
    p_x+\iim p_y & -p_z
    \end{pmatrix}
    \\
    p_{x}\pm\iim p_{y} &=
    -\iim \eem ^{\pm\iim \phi}
    \left(
    \frac{\partial}{\partial r}
    \pm\frac{\iim }{r}\frac{\partial}{\partial\phi}
    \right)
    \\
    p_{z} &=  -\iim \frac{\partial}{\partial z}
    \, ,
\end{align}
and the $z$ component of the orbital angular momentum operator is
\begin{align}
    \ell_z= -\iim \frac{\partial}{\partial \phi}  \, .
\end{align}
The volume element of integration transforms as
\begin{align}
    \int \dd^3 \bx = \int_0^\infty r \dd r \int_0^{2\pi} \dd\phi \int_{-\infty}^{\infty} \dd z \, .
    \label{eq:measure}
\end{align}
We align the molecule along the $z$ axis of the Cartesian coordinate system with the two nuclei placed at an equal distance from the origin. With that, the Coulomb potential takes the form
\begin{align}
    V(r,z) =  
    -\frac{Z_{1}}{\left(r^2 + (z+R/2)^2\right)^{1/2}}-\frac{Z_{2}}{\left(r^2 + (z-R/2)^2\right)^{1/2}}
    \, ,
\end{align}
where $R$ is the internuclear distance. The axial symmetry allows us to construct the wave function as an exact eigenfunction of the $j_z$ angular momentum operator, i.e. $j_z \psi_{a,m_j} = m_j \psi_{a,m_j}$, which implies 
\begin{align}
    \psi_{a,m_j}(r,\phi,z) & =
    \begin{pmatrix}
        \eem ^{\iim (m_j-1/2)\phi}f_1(r,z)\\
        \eem ^{\iim (m_j+1/2)\phi}f_2(r,z)\\
        \eem ^{\iim (m_j-1/2)\phi}f_3(r,z)\\
        \eem ^{\iim (m_j+1/2)\phi}f_4(r,z)
    \end{pmatrix}
    \, .
\end{align}
In the case of a homonuclear diatomic molecule, $Z_1=Z_2$, the wave function is also eigenfunction of the parity operator $\mathcal{P}$
\begin{align}
    \mathcal{P} \psi_{a,m_j}(\bx) = \beta \psi_{a,m_j}(-\bx) = \lambda \psi_{a,m_j}(\bx) \, ,
\end{align}
with $\lambda=+1$ or $\lambda=-1$, referred to as gerade and ungerade, respectively. The action of $\mathcal{P}$ on the wave function Ansatz is
\begin{align}
    \mathcal{P}    
    \psi_{a,m_j}(r,\phi,z)
    & =
    \begin{pmatrix}
     (-1)^{(m_j-1/2)}\eem ^{\iim (m_j-1/2)\phi}f_1(r,-z)\\
     (-1)^{(m_j+1/2)}\eem ^{\iim (m_j+1/2)\phi}f_2(r,-z)\\
    -(-1)^{(m_j-1/2)}\eem ^{\iim (m_j-1/2)\phi}f_3(r,-z)\\
    -(-1)^{(m_j+1/2)}\eem ^{\iim (m_j+1/2)\phi}f_4(r,-z)
    \end{pmatrix}
    \, ,
    \label{eq:parityop}
\end{align}
from which the transformation rule for each $f_i$ component of the wave function under $z\rightarrow -z$ can be deduced and the projection to the selected parity eigenstate can be implemented. 

The $f_i$ wave function components are expanded over a set of basis functions 
\begin{align}
    f_i(r,z) =\sum_\mu^{\nb} c^{(i)}_\mu
     g_\mu^{(i)}(r,z) \, ,
\end{align} 
where $g^{(i)}_\mu$ are chosen as products of two Gaussian functions, centered on each of the nuclei, projected to the appropriate parity, with a radial prefactor
\begin{align}
    g_\mu^{\pm}(r,z)
    =\mathcal{N}^{\pm}_{\mu} r^{|m_j\pm 1/2|} 
    \left[
    \eem^{-\zeta_{1,\mu}(r^2+(z+R/2)^2)-\zeta_{2,\mu}(r^2+(z-R/2)^2)} 
    +s_i(-1)^{m_j\pm 1/2}
    (\zeta_{1,\mu} \leftrightarrow \zeta_{2,\mu} ) 
    \right]
    \label{eq:gmu}
\end{align}
The function with the negative sign, $g_\mu^-$, is used for components $i=1,3$ while the function with the positive sign, $g_\mu^+$, is used for $i=2,4$. As a product of two Gaussians, this function can also be viewed as a single Gaussian `floating' along the $z$ axis. The $s_i=\pm 1$ factor is determined by the selected parity eigenstate. The normalization factor $\mathcal{N}_\mu^\pm$ is chosen such that
\begin{align}
    \braket{g_\mu^{\pm}}{g_\mu^{\pm}}_{r,z} 
    =
    1 \, ,
\end{align}
where the integral is understood over $r$ and $z$, with the integration measure given in Eq.~\eqref{eq:measure}. 

The upper ($i=1,2$) and lower ($i=3,4$) spinor components of the wave function are referred to as large and small components, respectively. To obtain a correct representation, it is necessary to consider the relation between the large and small components. We impose that the small-component basis functions are related to the large-component ones by the restricted kinetic balance (RKB) condition~\cite{StHa84}
\begin{align}
    \begin{pmatrix}
        g_\mu^{(3)}\\
        g_\mu^{(4)}
    \end{pmatrix}
        =
        \frac{\bsigma \cdot \bp }{2mc}
    \begin{pmatrix}
        g_\mu^{(1)}\\
        g_\mu^{(2)}
    \end{pmatrix} \, .
\end{align}
This relation also ensures that the small component has opposite parity from the large component (cf. Eq.~\eqref{eq:parityop}), therefore the projection onto the parity eigenstate is the same in the large and small component $f_i$ functions. Consequently, in Eq.~\eqref{eq:gmu} we set $s_i=\lambda$ for all components. We note that the RKB prescription favors the description of the positive-energy states. A more equal treatment can be obtained from the so-called dual-kinetic-balance condition~\cite{ShTuYePlSo04}, which is not considered in the present work. 

The RKB relation is most conveniently implemented as a transformation of the Hamiltonian and identity operators. We introduce the operator
\begin{align}
    X
    =
    \begin{pmatrix}
        1_2 & 0_2 \\
        0_2 & \frac{\bsigma\cdot \bp}{2mc}
    \end{pmatrix}
    \label{eq:RRKB} \, .
\end{align}
Under this transformation, the Hamiltonian becomes \cite{Kutzelnigg84,Dyall94}
\begin{align}
    h_{\text{RKB}} = X^\dagger h_\text{D} X
    =& 
    \begin{pmatrix}V & \frac{p^{2}}{2m}\\
    \frac{p^{2}}{2m} & \frac{\boldsymbol{\sigma}\cdot\boldsymbol{p}V\boldsymbol{\sigma}\cdot\boldsymbol{p}}{(2mc)^{2}}-\frac{p^{2}}{2m}
\end{pmatrix}
\end{align}
Similarly, the identity operator transforms as
\begin{align}
    I_\text{RKB}
    = 
    X^\dagger I X
    =
    \begin{pmatrix}
        1_2 & 0_2 \\
        0_2 & \frac{p^2}{(2mc)^2}
    \end{pmatrix}
    \, .
\end{align}
The matrix elements of these operators between Gaussian basis functions can be evaluated analytically in closed form for arbitrary $m_j$ values. For the overlap and differential operators the required matrix elements can be obtained from the formulas
\begin{align}
    \mathcal{S}_0 
    &=
    \int\text{d}^{3}\boldsymbol{x}\, 
    r^{n}
    \eem^{-\zeta_{ab}r^{2} -\zeta_{a} (z+R/2)^2 -\zeta_{b} (z-R/2)^2 } 
    \nonumber \\
    &=
    \frac{\pi^{3/2}}{\zeta_{ab}^{(3+n)/2}}
    \Gamma(1+n/2)
    \eem^{-\frac{\zeta_a\zeta_b}{\zeta_{ab}} R^2}
    \\
    \mathcal{S}_1 
    &=
    \int\text{d}^{3}\boldsymbol{x}\, z
    r^{n}
    \eem^{-\zeta_{ab}r^{2} -\zeta_{a} (z+R/2)^2 -\zeta_{b} (z-R/2)^2 } 
    \nonumber \\
    &= \frac{R(\zeta_b-\zeta_a)}{2\zeta_{ab}} \mathcal{S}_0
    \\
    \mathcal{S}_2
    &=
    \int\text{d}^{3}\boldsymbol{x}\, z^2
    r^{n}
    \eem^{-\zeta_{ab}r^{2} -\zeta_{a} (z+R/2)^2 -\zeta_{b} (z-R/2)^2 } 
    \nonumber\\
    &= \frac{R^2(\zeta_a-\zeta_b)^2 + 2\zeta_{ab}}{4\zeta_{ab}^2} \mathcal{S}_0 \, ,
\end{align}
where $\zeta_{ab}=\zeta_a+\zeta_b$. For the integrals with the Coulomb potential of one of the nuclei the following formulas are useful
\begin{align}
    \mathcal{V}_0
    &=\int\text{d}^{3}\boldsymbol{x}\,
    r^{n} 
    \left(r^{2}+\left(z+R/2\right)^{2}\right)^{-1/2}
    \eem^{-\zeta_{ab}r^{2} -\zeta_{a} (z+R/2)^2 -\zeta_{b} (z-R/2)^2 }
    \nonumber\\
    &= 
    \frac{\Gamma(n/2+1)}{\Gamma(n/2+3/2)}
    \zeta_{ab}^{1/2}
    \,_{1}F_{1}\left(1/2,(3+n)/2,-\frac{R^{2}\zeta_{b}^{2}}{\zeta_{ab}}\right) \mathcal{S}_0 
    \\
    \mathcal{V}_1
    &=\int\text{d}^{3}\boldsymbol{x}\, z
    r^{n} 
    \left(r^{2}+\left(z+R/2\right)^{2}\right)^{-1/2}
    \eem^{-\zeta_{ab}r^{2} -\zeta_{a} (z+R/2)^2 -\zeta_{b} (z-R/2)^2 }
    \nonumber\\
    &=
    -\frac{R\zeta_{ab}^{-1/2}
    \Gamma ( n / 2+1) }{2} 
    \left[
    \frac{\zeta_{ab}}{\Gamma ((3+n)/2)}
    \ _1F_1\left(
    1/2,
    (3+n)/2,
    -\frac{R^{2}\zeta_{b}^{2}}{\zeta_{ab}}
    \right) \right. \nonumber \\
    &\quad
    -
    \left. \frac{\zeta_b (2+n)}{\Gamma((5+n)/2)}
    \ _1F_1\left(
    1/2,
    (5+n)/2,
    -\frac{R^{2}\zeta_{b}^{2}}{\zeta_{ab}}
    \right)
    \right] \mathcal{S}_0
    \\
    \mathcal{V}_2
    &=\int\text{d}^{3}\boldsymbol{x}\, z^2
    r^{n} 
    \left(r^{2}+\left(z+R/2\right)^{2}\right)^{-1/2}
    \eem^{-\zeta_{ab}r^{2} -\zeta_{a} (z+R/2)^2 -\zeta_{b} (z-R/2)^2 }
    \nonumber\\
    &=
    \frac{\zeta_{ab}^{-3/2}
    \Gamma ( n / 2+1)}{4} 
    \left[
    \frac{\zeta_{ab}\left(  4+R^2\zeta_{ab} +2n \right)}{\Gamma ((3+n)/2)}
    \ _1F_1\left(
    1/2,
    (3+n)/2,
    -\frac{R^{2}\zeta_{b}^{2}}{\zeta_{ab}}
    \right) \right. \nonumber \\
    &\quad 
    -
    \left. \frac{(2+n) 
    \left(
    \zeta_{ab}\left(2+2R^2\zeta_b+n\right)
    -2R^2\zeta_b^2
    \right)
    }{\Gamma((5+n)/2)}
    \ _1F_1\left(
    1/2,
    (5+n)/2,
    -\frac{R^{2}\zeta_{b}^{2}}{\zeta_{ab}}
    \right)
    \right]  \mathcal{S}_0  \, .
\end{align}
The corresponding integrals with the potential of the other nucleus are obtained by exchanging $\zeta_a \leftrightarrow \zeta_b$. 

The basis set parameters were generated analogously to the stochastic variational method~\cite{SuVa98}. We start with at least one pair of exponents, which can be chosen randomly, provided that their sum is positive, so that the functions are square integrable. A new pair of exponents is generated by randomly selecting one of the existing pairs and multiplying it by a random positive real number. Several trial functions are generated in this way and the nonrelativistic energy is evaluated by adding each of them to the current basis set. The exponent pair that yields the lowest nonrelativistic energy is kept. This simple approach is sufficiently robust and allows for the systematic enlargement of the basis set. Upon the selection of several new exponents, the parameters are optimized globally using the \textsc{bobyqa} algorithm~\cite{bobyqa} to minimize the nonrelativistic energy. To enforce the condition that the sum of the two exponents remains positive $\zeta_{1,\mu}+\zeta_{2,\mu} >0$, the optimization is more conveniently carried out in terms of the combinations
\begin{align}
    \begin{split}
        \zeta_\mu^+ &= \zeta_{1,\mu} + \zeta_{2,\mu} \in (0,\zeta_\text{max}) \\
        \zeta_\mu^- &=| \zeta_{1,\mu} - \zeta_{2,\mu} | \in (0,\zeta_\text{max}) \, ,
    \end{split}
\end{align}
where $\zeta_\text{max}$ was chosen as $10^{15}$ in this work. The same basis set, optimized for the nonrelativistic energy, was used to evaluate the relativistic energies by diagonalization of the Dirac--Coulomb Hamiltonian matrix. This procedure yields satisfactory results for low-$Z$ systems, where the nonrelativistic wave function is a reasonable approximation.

The exact Dirac--Coulomb wave function has a weak singularity in the vicinity of point-like nuclei, with the short-distance behavior $r^{\gamma-1}$, where $\gamma = \sqrt{1-(Z\alpha)^2}$ for the ground state \cite{DyallFaegri07}. This nonanalytic behavior cannot be represented exactly by the present Ansatz and requires large exponents in the basis set. The nonrelativistic wave function has a cusp at the nuclei, which also requires large exponents. Consequently, even though the basis-set parameters are optimized with respect to the nonrelativistic energy, these large exponents appear in the optimized basis set (see Supplementary Material) and allow for the approximate description of the weak singularity in the relativistic case. We note that in the case of an extended nuclear-charge distribution these singularities are absent and the Gaussian basis correctly represents the short-range behavior of the wave function \cite{IshikawaQuiney87}.

\subsection{Momentum-space wave function}
A significant advantage of Gaussian-type functions is that the Fourier transform is analytically known and has a simple form. For our Ansatz the momentum-space wave function takes the form
\begin{align}
    \psi_{a,m_j}(p_\perp,\chi,p_z) 
    =
    \int 
    \frac{\dd \bx}{(2\pi)^{3/2}}
    \eem^{-\iim \bp \cdot \bx}
    \psi_{a,m_j}(\bx) 
    =
    \begin{pmatrix}
        \eem ^{\iim (m_j-1/2)\chi}\tilde{f}_1(p_\perp,p_z)\\
        \eem ^{\iim (m_j+1/2)\chi}\tilde{f}_2(p_\perp,p_z)\\
        \eem ^{\iim (m_j-1/2)\chi}\tilde{f}_3(p_\perp,p_z)\\
        \eem ^{\iim (m_j+1/2)\chi}\tilde{f}_4(p_\perp,p_z)
    \end{pmatrix}
\end{align}
where $\{p_\perp, \chi, p_z \}$ denote the momentum-space cylindrical coordinates. The $\tilde{f}_i(p_\perp,p_z)$ functions are expanded over the Fourier transform of the Gaussian basis functions
\begin{align}
    \tilde{g}_\mu^{\pm}(p_\perp,p_z)
    =
    \tilde{\mathcal{N}}^{\pm}_{\mu} p_\perp^{|m_j\pm 1/2|} 
    \left[
    \eem^{- \frac{1}{4\zeta_{12,\mu}}  
    \left[
    p^2 - 2\iim R p_z (\zeta_{1,\mu}-\zeta_{2,\mu})
    \right] }
    +s_i(-1)^{m_j\pm 1/2}
    (\zeta_{1,\mu} \leftrightarrow \zeta_{2,\mu} ) 
    \right]
\end{align}
with $p^2=p_\perp^2+p_z^2$ and $\zeta_{12,\mu}=\zeta_{1,\mu}+\zeta_{2,\mu}$. These wave functions are useful for the evaluation of expectation values involving functions of the momentum operator, as in the case for instance in the computation of the self-energy correction.

The many-potential-expansion method for the computation of the bound-electron self-energy is described in detail in Refs.~\cite{Snyderman91,BlundellSnyderman91,YerokhinShabaev99}. In this framework the renormalized one-loop self-energy correction is given as the sum of three terms
\begin{align}
    \Delta E^\text{SE} 
    = 
    \Delta E^{(0)}
    +
    \Delta E^{(1)}
    +
    \Delta E^\text{mp} 
    \, .
    \label{eq:mpe}
\end{align}
The first two terms, the zero- and one-potential terms, are computed in momentum space, using the Fourier transform of the wave function, while the last term, the many-potential term, involves an infinite sum over intermediate states, and it is evaluated in coordinate space. The many-potential term is the most complicated of all, and it is not considered here. As for the momentum-space contributions, there is no conceptual difference between the zero- and one-potential terms, as they are both obtained from the Fourier transform of the wave function, yet the computation of the latter is technically more involved. Therefore, as a proof of concept, we will focus here only on the zero-potential term, that is, the expectation value of the renormalized free-electron self-energy operator
\begin{align}
    \Delta E_a^{(0)} 
    = 
    \int \dd \bp \, \bar{\psi}_a(\bp)
    \Sigma_\text{R}(E_a,p) \psi_a(\bp)
    \, ,
\end{align}
where $\bar{\psi} = \psi^\dagger \beta$ is the Dirac adjoint. When evaluating these terms, it is more convenient to work in natural units, where $c=1$. Furthermore, the energy $E_a$ should be considered \emph{without} the $-mc^2$ shift introduced in the Hamiltonian.

The free-electron self-energy can be derived in different gauges of the photon propagator. The most common choice is Feynman gauge for its simplicity, which takes the form \cite{YerokhinShabaev99}
\begin{align}
    \Sigma^{(0)}_\text{R,F}(p_0,p)
    =
    \frac{\alpha}{4\pi} 
    \left[
    a(\rho)
    +
    (\gamma^0 p_0 - \bos{\gamma}\cdot \bp)
    b(\rho)
    \right]
    \, ,
\end{align}
with the definition of the coefficients given in Appendix~\ref{app:FSE}, Eqs~\eqref{eq:SEcoeffFr}-\eqref{eq:SEcoeffFb}. As it was pointed out already in Ref.~\cite{Snyderman91}, the individual terms of the many-potential expansion contain spurious contributions in Feynman gauge of order $\alpha(Z\alpha)^2$, which cancel in the final result. The large cancellations among the individual contributions make the numerical evaluation for low nuclear charges ($Z \leqslant 5$) cumbersome. This spurious contribution, however, is absent in Fried--Yennie, or in Coulomb gauges~\cite{Snyderman91,HeHo12,YeHaKe25}. As the use of Coulomb gauge was found particularly useful for low nuclear charges~\cite{YeHaKe25}, we carry out the computation also in Coulomb gauge, where the free-electron self-energy operator takes the form~\cite{Adkins83}
\begin{align}
    \Sigma^{(0)}_\text{R,C}(p_0,p)
    =
    \frac{\alpha}{4\pi} 
    \left[
    \tilde{a}(p_0,p)
    +p_0\gamma^0 \tilde{b}(p_0,p)
    +\boldsymbol{\gamma} \cdot \bp 
    \tilde{c}(p_0,p)
    \right]
\end{align}
The explicit form of the coefficients is given in Appendix~\ref{app:FSE}, Eqs.~\eqref{eq:SEcoeffCa}--\eqref{eq:SEcoeffCc}. 

When evaluating the zero-potential term it is convenient to transform the operators according to the RKB prescription, analogously to the Hamiltonian as
\begin{align}
    \Sigma_\text{R} 
    \rightarrow 
    X^\dagger \Sigma_\text{R} X
    \, ,
\end{align}
where $X$ is defined in Eq.~\eqref{eq:RRKB}. This results in the following integrals for Feynman and Coulomb gauges, respectively
\begin{align}
    \Delta E_{a,\text{F}}^{(0)} 
    &= 
    \int \dd \bp \,
    \bar{\psi}_a(\bp) \Sigma^{(0)}_\text{R,F}(E_a,p) \psi_a(\bp)
    \nonumber \\
    &=\frac{\alpha}{2} 
    \int_0^\infty \dd p_\perp\, p_\perp \int_{-\infty}^\infty \dd p_z \,
    \left[
    \left(a(\rho)+b(\rho) p^0\right) \left(|\tilde{f}_1|^2 + |\tilde{f}_2|^2\right)
    \right.
    \nonumber \\
    &\quad 
    - \frac{p^2}{mc}b(\rho) \mathfrak{Re}\left(\tilde{f}_1^\ast \tilde{f}_3 + \tilde{f}_2^\ast \tilde{f}_4\right) 
    \left.
    + \left(-a(\rho) + b(\rho)p^0 \right) 
    \frac{p^2}{(2mc)^2}  \left(|\tilde{f}_3|^2 + |\tilde{f}_4|^2\right)
    \right]
    \label{eq:E0F}
\end{align}
\begin{align}
    \Delta E_{a,\text{C}}^{(0)} 
    &= 
    \int \dd \bp \,
    \bar{\psi}_a(\bp) \Sigma^{(0)}_\text{R,C}(E_a,p) \psi_a(\bp)
    \nonumber \\
    &=\frac{\alpha}{2} 
    \int_0^\infty \dd p_\perp\, p_\perp \int_{-\infty}^\infty \dd p_z \,
    \left[
    \left(\tilde{a}(E_a,p)+\tilde{b}(E_a,p) E_a\right) \left(|\tilde{f}_1|^2 + |\tilde{f}_2|^2\right)
    \right.
    \nonumber \\
    & \quad 
    + \frac{p^2}{mc}\tilde{c}(E_a,p) \mathfrak{Re}\left(\tilde{f}_1^\ast \tilde{f}_3 + \tilde{f}_2^\ast \tilde{f}_4\right) \left.
    + \left(-\tilde{a}(E_a,p) + \tilde{b}(E_a,p) E_a \right) 
    \frac{p^2}{(2mc)^2}  \left(|\tilde{f}_3|^2 + |\tilde{f}_4|^2\right)
    \right]
    \label{eq:E0C}
    \, .
\end{align}
The remaining integrals over the momentum variables can be done numerically. Analogously to atomic computations, we use the dimensionless quantity $F(Z\alpha)$, defined as
\begin{align}
    \Delta E^\text{SE} 
    = 
    \frac{\alpha}{\pi} (Z\alpha)^4 F(Z\alpha) m c^2 \, ,
    \label{eq:FZa}
\end{align}
to present the results in a form that is consistent with previous works, here with $Z=Z_1=Z_2=1$. We note that, in the molecular case, the function $F(Z\alpha)$ also depends parametrically on the nuclear distance and the electronic state. Finally, as only the zero-potential term is computed in the present work, the results given in the following section in terms of $F(Z\alpha)$ correspond to the contribution $\Delta E^{(0)}$ in Eq.~\eqref{eq:mpe}.

\section{Results and discussion}
\begin{figure}
    \centering
    \includegraphics[width=0.7\linewidth]{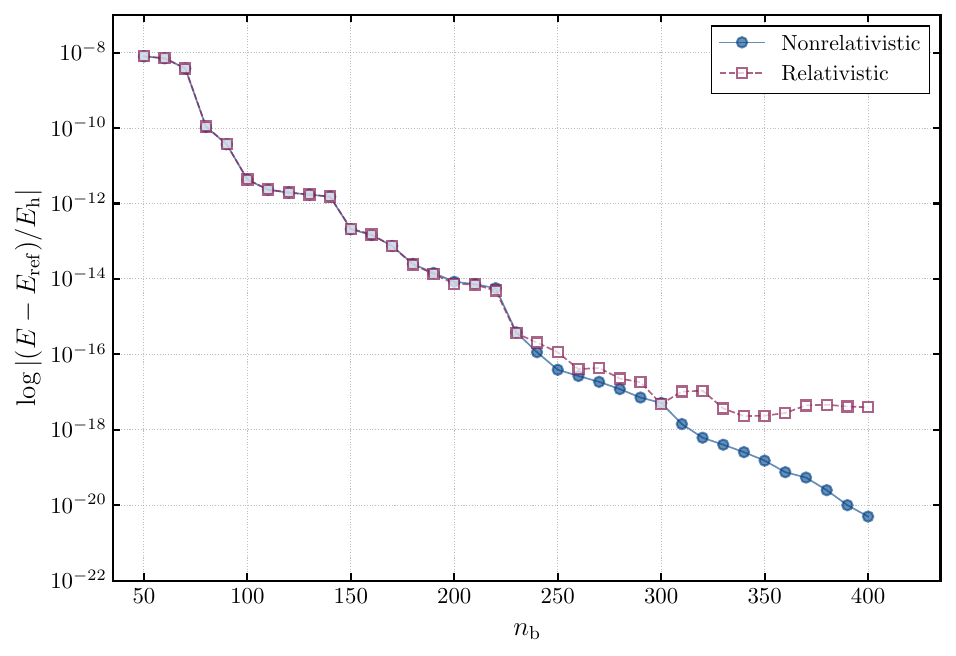}
    \caption{Convergence with the size of the basis set of the relativistic and nonrelativistic energies on a log scale, with respect to reference values, $E_\text{ref}^\text{nr}=-1.102\, 634\, 214\, 494\, 946\, 461\, 508\, 98$ and $E_\text{ref}^\text{rel}= -1.102\, 641\, 581\, 032\, 577\, 164\, 118\, 14 $ of Ref.~\cite{Kullie24}. }
    \label{fig:convergence}
\end{figure}

The method was implemented in Fortran 90, and all computations were carried out using quadruple precision arithmetic. The convergence of the nonrelativistic and relativistic energy with the basis set is shown in Table~\ref{tab:energy}. The nonrelativistic energy converges monotonically, reaching a $10^{-20}$ level relative precision. The relativistic energy, upon reaching a relative precision of $10^{-18}$, does not improve upon the enlargement of the basis set. This is presumably due to the limitation of the nonrelativistic optimization of the basis set parameters. The convergence is also shown in Figure~\ref{fig:convergence} on a logarithmic scale. The convergence of the nonrelativistic energy is close to linear (on logarithmic scale), while the convergence of the relativistic energy first follows that of the nonrelativistic one, then reaches a plateau. Further improvement of the relativistic energy would require the optimization of the parameters with respect to the relativistic energy. We note that no numerical instability associated with linear dependence was observed at quadruple precision. The basis set parameters are given in Supplementary Material. 

As a first application of the momentum-space wave function, we evaluate the zero-potential contribution to the one-loop self-energy. We emphasize that this quantity is only a partial contribution to the complete bound-electron self-energy and it is not separately gauge invariant. Nevertheless, its evaluation demonstrates the applicability of the analytic Fourier transform of the wave function to momentum-space expectation values. The one-potential term, although technically more involved, can in principle be evaluated along similar lines, using the momentum-space wave function. The final integrals over the remaining momentum variables in Eqs.~\eqref{eq:E0F} and \eqref{eq:E0C} were carried out using the \textsc{cuhre} subroutine of the \textsc{cuba} library, compiled with 128 bit arithmetic real variables (quadruple precision)~\cite{Hahn07}. The results in both gauges for various basis sets are shown in Table~\ref{tab:zeropot}. The quantity converges rapidly with the basis set size in both cases. As expected from the atomic case, the magnitude of the term in Feynman gauge is four orders of magnitude larger than in Coulomb gauge. 

To check the correctness of the implementation, we have repeated the computation for $R=0$, using an even-tempered basis set for one of the exponents $\zeta_{1,\mu}= \zeta_{1,1} \beta^{i-1}$, where $\ i=1,\ldots \nb$ with parameters $\zeta_{1,1}=10^{-4}~a_0^{-2}$, $\beta = 1.3$ and $\nb = 100$. The other exponent was set to zero and the parity projection was turned off. In the united-atom limit, the two-center Coulomb potential reduces to the central Coulomb potential of a nucleus with charge $Z=2$. The above basis set reproduces the exact relativistic energy to approximately 12 significant digits. For the zero-potential term we obtain the following result, using $Z=2$ in Eq.~\eqref{eq:FZa} to convert to dimensionless units
\begin{align}
    \Delta E^{(0)}_\text{F}(R=0) =
    -35\, 550.985\, 179
    &&
    \Delta E^{(0)}_\text{C}(R=0) =
    11.181\, 951\, 6
    \, ,
\end{align}
in excellent agreement to all given digits with the zero-potential contribution for He$^+$, computed with the exact Dirac--Coulomb wave function. At large internuclear distances, the results should approach the hydrogen atom limit. We therefore studied the $R\rightarrow \infty$ asymptotic behavior, by repeating the computation for $R=50,\ 100,\ 150$, and $ R=200$ bohr nuclear distances, using the same, even-tempered basis set as above, but with projection to the gerade eigenstate. For the Feynman-gauge results a slightly larger $\nb=150$ basis set was used to converge all given significant digits. The results were extrapolated by fitting a polynomial of the form $P(R^{-1})=c_0 + c_1R^{-1} + c_2R^{-2} + c_3R^{-3}$. The fitted value of the $c_0$ coefficient recovers the exact result for the hydrogen atom. The obtained values for the various distances and the extrapolated values are shown in Table~\ref{tab:largeR}. Note, that the exact values differ slightly from the ones of Ref.~\cite{YeHaKe25}, due to the different numerical value of the fine-structure constant used in the present work. 

\begin{table}
    \centering
    \caption{Convergence of the nonrelativistic and relativistic energies with the stochastic optimization method at $R=2.00~a_0$ nuclear distance. All values in atomic units.}
    \label{tab:energy}
        \begin{tabular}{@{} c@{\ \ } d{2.30}@{\ } d{2.30}  @{}}
        \hline \hline
         \multicolumn{1}{c}{$\nb$}& \multicolumn{1}{c}{$E_\text{nr}$} 
         & \multicolumn{1}{c}{$ E_\text{rel}$}
          \\
        \hline
        50  & -1.102\, 634\, 206\, 491\, 406\, 987\, 688  & -1.102\, 641\, 573\, 028\, 653\, 782\, 043 \\
        100 & -1.102\, 634\, 214\, 490\, 621\, 503\, 582  & -1.102\, 641\, 581\, 028\, 241\, 986\, 391 \\
        150 & -1.102\, 634\, 214\, 494\, 737\, 576\, 317  & -1.102\, 641\, 581\, 032\, 362\, 262\, 938 \\
        200 & -1.102\, 634\, 214\, 494\, 938\, 063\, 257  & -1.102\, 641\, 581\, 032\, 569\, 668\, 770 \\
        250 & -1.102\, 634\, 214\, 494\, 946\, 422\, 445  & -1.102\, 641\, 581\, 032\, 577\, 050\, 456 \\
        300 & -1.102\, 634\, 214\, 494\, 946\, 456\, 347  & -1.102\, 641\, 581\, 032\, 577\, 168\, 956 \\
        350 & -1.102\, 634\, 214\, 494\, 946\, 461\, 357  & -1.102\, 641\, 581\, 032\, 577\, 166\, 476 \\
        400 & -1.102\, 634\, 214\, 494\, 946\, 461\, 504  & -1.102\, 641\, 581\, 032\, 577\, 168\, 053 \\
        \hline 
        Ref.~\cite{Kullie24,NogueiraKarr23} & 
        -1.102\, 634\, 214\, 494\, 946\, 461\, 508\, 98  &
        -1.102\, 641\, 581\, 032\, 577\, 164\, 118\, 14     \\
        \hline \hline
        \end{tabular}
\end{table}
\begin{table}
    \centering
    \caption{Convergence of the zero-potential term in Coulomb and Feynman gauge at $R=2.00~a_0$ internuclear distance expressed in terms of $F(Z\alpha)$, defined in Eq.~\eqref{eq:FZa}.
    }
    \label{tab:zeropot}
        \begin{tabular}{@{} c@{\ } d{3.10}@{\ \ \ } d{8.11}  @{}}
        \hline \hline
        \multicolumn{1}{c}{$n_\text{b}$}
        & 
        \multicolumn{1}{c}{
        $\Delta E^{(0)}_\text{C}$
        } 
        & 
        \multicolumn{1}{c}{
        $\Delta E^{(0)}_\text{F}$
        }
        \\
        \hline
        50  & 20.541\, 435\, 280  & -272\, 863.830\, 452\, 684 \\
        100 & 20.541\, 435\, 117  & -272\, 863.830\, 442\, 730 \\
        150 & 20.541\, 435\, 135  & -272\, 863.830\, 443\, 948 \\
        200 & 20.541\, 435\, 136  & -272\, 863.830\, 444\, 041 \\
        250 & 20.541\, 435\, 135  & -272\, 863.830\, 444\, 042 \\
        \hline \hline
        \end{tabular}
\end{table}
\begin{table}
    \centering
    \caption{Large-$R$ asymptotic behavior of the zero-potential term, with extrapolation to infinite nuclear distance, compared to the exact hydrogen value. Units are $F(Z\alpha)$, defined in Eq.~\eqref{eq:FZa}.}
    \label{tab:largeR}
        \begin{tabular}{@{} c@{\ } d{3.9}@{\ \ \ } d{8.7}  @{}}
        \hline \hline
        \multicolumn{1}{c}{$R$}
        & 
        \multicolumn{1}{c}{
        $\Delta E^{(0)}_\text{C}$
        } 
        & 
        \multicolumn{1}{c}{
        $\Delta E^{(0)}_\text{F}$
        }
        \\
        \hline
        50  & 13.940\, 954  & -171\, 284.632\, 40 \\
        100 & 13.895\, 263  & -169\, 731.549\, 34 \\
        150 & 13.880\, 012  & -169\, 213.342\, 67  \\
        200 & 13.872\, 381  & -168\, 954.142\, 61 \\
        \text{Extrap.} &
        13.849\, 474(1) & -168\, 176.153\, 7(3) \\
        \hline 
        \text{Exact} &  13.849\, 474\, 032 & -168\, 176.153\, 752 \\
        \hline \hline
        \end{tabular}
\end{table}

\section{Conclusion and outlook}
We have presented a Gaussian-basis-set approach for the numerical solution of the two-center Dirac equation. The Gaussian-product Ansatz, adapted to the axial symmetry of diatomic molecules, retains most of the analytic properties of radial Gaussians and at the same time it provides energies with relative precision at the $10^{-18}$ level. Although the numerical precision obtained for the ground-state energy is below that achieved in the highest-precision calculations with exponential-type basis sets, the present approach yields sufficiently accurate relativistic energies with a compact basis set, and it is numerically robust with quadruple-precision arithmetic. 
We exploited the analytic Fourier transform of the basis functions to evaluate the zero-potential contribution to the one-loop self-energy for a molecular ion in both Feynman and Coulomb gauges. The results also reproduce the expected qualitative difference between the two gauges, the absence of the spurious lower-order contribution in Coulomb gauge. The present approach provides a route toward the evaluation of the remaining contributions to the one-loop self-energy for hydrogen molecular ions, which will be considered in future work.

\begin{acknowledgments}
I thank Trond Saue for his continuous support, motivating discussions, and suggestions on the manuscript. I also thank Jean-Philippe Karr for the discussions, and Edit M\'atyus for the quadruple-precision linear algebra subroutines. This project was funded by the European Research Council (ERC) under the European Union’s Horizon 2020 research and innovation programme (Grant Agreement No. ID:101019907).
\end{acknowledgments}

\appendix
\section{Free-electron self-energy operator}

The formulas are given in relativistic units ($\hbar=c=m=1$). The coefficients appearing in the free-electron self-energy operator in Feynman gauge are~\cite{Snyderman91,YerokhinShabaev99}
\label{app:FSE}
\begin{align}
    \rho 
    &= 1- p_0^2 + p^2
    \label{eq:SEcoeffFr}
    \\
    a(\rho)
    &= 2\left(
    1+\frac{2\rho}{1-\rho} \ln\rho
    \right) 
    \label{eq:SEcoeffFa}
    \\
    b(\rho) 
    &= -\frac{2-\rho}{1-\rho}
    \left(
    1+\frac{\rho}{1-\rho} \ln\rho
    \right)
    \label{eq:SEcoeffFb}
\end{align}

The expression in Coulomb gauge was first given in terms of Feynman parameter integrals~\cite{Adkins83}. This integrals can be solved either numerically as done in Ref.~\cite{YeHaKe25}, or analytically in terms of the dilogarithm function~\cite{Malenfant87}. In the present work, since the integral over momentum is already two-dimensional, we use the expressions in the form
\begin{align}
    \tilde{a}(p_0,p)
    &=
    6-\frac{4\sqrt{1+p^2}}{p} \sinh^{-1}(p)
    -
    \frac{2(1+p^2-p_0^2)}{p^2-p_0^2} 
    \ln\left( 1+p^2-p_0^2 \right)
    \label{eq:SEcoeffCa}
    \\
    \tilde{b}(p_0,p)
    &=
    -\frac{1}{2}
    -
    \frac{p^2-p_0^2-2}{2(p^2-p_0^2)}
    +\frac{(p^2-p_0^2)^2-1}{(p^2-p_0^2)^2}
    \ln\left(
    1+p^2-p_0^2
    \right)
    \label{eq:SEcoeffCb}
    \\
    \tilde{c}(p_0,p)
    &=
    \frac{19}{6}
    -\frac{4(3-5p^2)}{9p^2}
    -\frac{4\sqrt{1+p^2}}{p^3} 
    \left(
    1+2(p^2-p_0^2)
    \right)
    \sinh^{-1}(p) 
    \nonumber \\
    &\quad
    +
    \frac{p^2-p_0^2-2}{2(p^2-p_0^2)}
    -\frac{(p^2-p_0^2)^2-1}{(p^2-p_0^2)^2}
    \ln\left(
    1+p^2-p_0^2
    \right) 
    \nonumber \\
    &\quad
    +\frac{4(12+7p^2-18p_0^2+9(1+p^2-p_0^2)\ln\left(1+p^2-p_0^2\right) )}{9p^2} 
    \nonumber\\
    &\quad
    +\frac{2p_0(p_0^2-p^2-1)}{p^3}
    \left[
    -\text{Li}_2 \left(
    \frac{p-p_0}{\sqrt{1+p^2}-p_0}
    \right)
    +\text{Li}_2 \left(
    -\frac{p_0+p}{\sqrt{1+p^2}-p_0}
    \right) \right.
    \nonumber\\
    &\quad
    \left.-\text{Li}_2 \left(
    \frac{p_0-p}{\sqrt{1+p^2}+p_0}
    \right)
    +\text{Li}_2 \left(
    \frac{p+p_0}{\sqrt{1+p^2}+p_0}
    \right)
    \right]    
    \label{eq:SEcoeffCc}
\end{align}

In both cases special care must be taken where the denominator vanishes. Near these points, the formula is evaluated through its Taylor series.

\newpage
\bibliography{references}
\end{document}